\documentclass[sigconf,screen]{acmart}
\AtBeginDocument{%
  \providecommand\BibTeX{{%
    \normalfont B\kern-0.5em{\scshape i\kern-0.25em b}\kern-0.8em\TeX}}}

\copyrightyear{2021} 
\acmYear{2021} 
\setcopyright{acmlicensed}\acmConference[ESEM '21]{ACM / IEEE International Symposium on Empirical Software Engineering and Measurement (ESEM)}{October 11--15, 2021}{Bari, Italy}
\acmBooktitle{ACM / IEEE International Symposium on Empirical Software Engineering and Measurement (ESEM) (ESEM '21), October 11--15, 2021, Bari, Italy}
\acmPrice{15.00}
\acmDOI{10.1145/3475716.3484191}
\acmISBN{978-1-4503-8665-4/21/10}

\usepackage{todonotes}
\usepackage{enumitem}
\usepackage{titlesec}
\titlespacing{\section}{0pt}{0.01cm}{0pt}
\titlespacing{\subsection}{0pt}{0.005cm}{0pt}
\titlespacing{\subsubsection}{0pt}{0.0025cm}{0pt}

\begin{document}

\title{Vision for an Artefact-based Approach to Regulatory Requirements Engineering}

\author{Oleksandr Kosenkov} 
\email{kosenkov@fortiss.org}
\affiliation{
    \institution{fortiss GmbH}
    \city{Munich}
    \country{Germany}}

\author{Michael Unterkalmsteiner} 
\email{michael.unterkalmsteiner@bth.se}
\affiliation{%
  \institution{Blekinge Institute of Technology}
  \department{Software Engineering Research and Education Lab}
  \city{Karlskrona}
  \country{Sweden}}

\author{Daniel Mendez} 
\email{daniel.mendez@bth.se}
\affiliation{%
  \institution{Blekinge Institute of Technology}
  \department{Software Engineering Research and Education Lab}
  \city{Karlskrona}
  \country{Sweden}}

\author{Davide Fucci}
\email{davide.fucci@bth.se}
\affiliation{%
  \institution{Blekinge Institute of Technology}
  \department{Software Engineering Research and Education Lab}
  \city{Karlskrona}
  \country{Sweden}}

\renewcommand{\shortauthors}{Kosenkov et al.}

\begin{abstract}
\emph{Background:} Nowadays, regulatory requirements engineering (regulatory RE) faces challenges of interdisciplinary nature that cannot be tackled due to existing research gaps.
\emph{Aims:} We envision an approach to solve some of the challenges related to the nature and complexity of regulatory requirements, the necessity for domain knowledge, and the involvement of legal experts in regulatory RE.
\emph{Method:} We suggest the qualitative analysis of regulatory texts combined with the further case study to develop an empirical foundation for our research.
\emph{Results:} We outline our vision for the application of extended artefact-based modeling for regulatory RE.
\emph{Conclusions:} Empirical methodology is an essential instrument to address interdisciplinarity and complexity in regulatory RE. Artefact-based modeling supported by empirical results can solve a particular set of problems while not limiting the application of other methods and tools and facilitating the interaction between different fields of practice and research.
\end{abstract}




\maketitle

\section{Introduction}\label{sec:introduction}
Nowadays, organizations worldwide prioritize digital transformation for future development~\cite{deloitte2021} and simultaneously face over-regulation and policy uncertainty~\cite{pwc2020}. These trends support the assumption that legal compliance can become one of the most important non-functional software requirements in the near future~\cite{massey2011}. Indeed, as software systems have become an ubiquitous and essential element of modern society and organizations, their design and application should conform to a wide range of regulations. Regulatory requirements engineering is one of the RE research areas~\cite{boella2014} contributing to the compliance of software systems by processing regulatory requirements for software engineering purposes. Regulatory requirements are, simply speaking, requirements that are derived from mandatory public norms contained in laws and other regulations. Such regulations have a particular purpose and even use a specific ``legal language in the service of regulating social behavior'' \cite{maley1994}. Regulatory norms---logical units of regulations containing a single regulatory statement---rarely target software systems directly but rather pursue more general regulatory purposes. The overall goal of regulatory norms and requirements contrasts with the goals of software requirements which capture user requirements and technical constraints that need to be met by a software system. This difference is the source of fundamental challenges in regulatory RE, which we describe next.

\subsection{The wickedness of regulatory RE}\label{sec:wicked}
First and foremost, regulatory requirements are different from other requirements and need special engineering methods because: (1) a multiplicity of regulators as stakeholders issue relevant regulatory norms and a variety of forms of such regulatory norms~\cite{otto2009, ghanavati2014a}; (2) the nature of regulatory norms as primary source of regulatory requirements. Often, regulatory norms are expressed for organizations as a whole~\cite{compagna2007}, and the identification of applicable regulatory norms requires business, organization, and community context~\cite{otto2009}; (3) the formulation of regulatory norms is abstract~\cite{breaux2007}, ambiguous, and vague~\cite{massey2015}; (4) the changeability of regulatory requirements due to variability, and evolution of regulations over time~\cite{maxwell2012}.

Second, the interpretation of regulatory requirements typically needs the legal knowledge of legal experts. Regulatory RE research faces the challenge of capturing tacit legal knowledge~\cite{jureta2013} to address the peculiarities of regulatory requirements described above. In practice, the interdisciplinary nature of regulatory RE makes it important to establish a dialog between legal experts and requirements engineers for an effective regulatory RE process~\cite{boella2014}, involve legal experts when legal expertise is required~\cite{massey2015}, and ensure that the legal validation of a software system can be conducted by lawyers~\cite{bobkowska2010}. As a result, it is necessary to ensure clear communication between different stakeholders when it comes to regulatory compliance, and hence guarantee a ``sufficient level of acceptability'' from the point of view of legal practitioners~\cite{boella2014}.

\subsection{Our vision of regulatory RE}
In this paper, we present our vision of the application of the Artefact Model for Domain-independent Regulatory Requirements Engineering (AMDiRRE)~\ref{fig:amdirre} and legal interpretation methodological framework. Their main purposes are: (1) resolve the peculiarities of regulatory requirements that cannot be addressed with existing RE approaches; (2) address the practical regulatory RE and compliance needs, primarily through enabling the interaction between requirements engineers and legal experts, (3) suggest approaches for incorporating empirical software engineering methodologies into regulatory RE research and practice.

The rest of the paper is structured as follows. In Section~\ref{sec:gaps} we describe potential key gaps in past regulatory RE research. Section~\ref{sec:concept} describes the concept of our approach. We share our vision for the regulatory RE domain in the coming decade and outline the potential contribution of our approach in Section~\ref{sec:vision}. Section~\ref{sec:conclusion} concludes the paper.

\section{Gaps in Regulatory RE Research}\label{sec:gaps}
Regulatory RE research has, so far, primarily focused on addressing different features of regulatory norms that distinguish them from other sources of requirements (see Section~\ref{sec:wicked}). Regulatory RE methods and tools address the peculiarities of regulatory norms and requirements in an isolated manner, concentrating on topmost domains such as healthcare or finance~\cite{akhigbe2019}. There is a lack of methods that systematically cover all facets of regulatory compliance problems~\cite{hashmi2018, akhigbe2019}. The lack of a systematic approach in regulatory RE has resulted in naive, and relatively simplistic views of regulations in comparison to the legal perspective~\cite{boella2014}. As a result, some of the approaches that cannot capture important facets of regulations are criticized by legal experts~\cite{boella2014}. We conjecture that the fragmented character of regulatory RE approaches constitutes a barrier to the incorporation of legal knowledge. Legal and software engineering perspectives on the implementation of regulatory requirements have a number of differences in some requirements' aspects (e.g., semantics or scope of requirements)~\cite{kosenkov2021} that lead to discontinuity if not considered systematically. In this section, we describe the research gaps that, based on our analysis, led to the absence of systematicity in regulatory RE methods.

\subsection{Lack of integrated approaches}\label{sec:non-integrated}
The elicitation of the meaning of regulatory norms and the analysis of regulatory requirements is the main focus of regulatory RE research. However, the aspects related to system and structure of regulations such as multiplicity of regulations~\cite{ghanavati2014a}, different types of regulatory sources~\cite{otto2009}, or cross-references~\cite{sannier2017} have received less attention. Only a few methods have addressed both system (interconnections between different regulations) and structure of law in combination with semantics (e.g.,~\cite{ghanavati2014b}). Nevertheless, even when system and structure of regulations are addressed along with semantics, their interrelations are not sufficiently investigated. It remains unclear how exactly system and structure of law influence the semantics---e.g., Ghanavati et al.~\cite{ghanavati2014b} models external cross-references as goals. Although the relevance of system and structure of law for the meaning of regulatory norms is recognized, regulatory RE tends to focus on the level of single norms in single regulations~\cite{boella2014}. At the same time, legal research recognizes that the structure of the legal system is relevant for inferring the meaning of rules~\cite{wroblewski1985}. For example, the legal domain can be classified into public law (i.e., the regulation of social relations that involves direct social concern) and private law (i.e., the regulations that involve social relations between private persons); moreover, it contains multiple fields of law (e.g., privacy law, media law). This separation leads to distinctive features that determine what kind of social relations the law addresses and what regulatory methods are applied, implying that the regulatory RE approaches need to consider the \textit{context} of the legal domain.

\subsection{Limitations in semantic analysis}
Existing regulatory RE methods for addressing regulatory norms and requirements at a semantic level do seldom consider the peculiar nature of legal norms. Semantic analysis methods (e.g.,~\cite{breaux2007, sleimi2021}) assume that it is possible to derive requirements from the regulations' immediate analysis (e.g., using an upper ontology~\cite{breaux2007} or semantic metadata~\cite{sleimi2021}).
However, legal scholarship provides a different perspective on the nature of regulatory norms and their meaning, as discussed next.

\subsubsection{Peculiar language}
Regulatory norms are specified using a special legal language, characteristics of which are related to the meaning of regulatory norms~\cite{wroblewski1985}. Such features include distinctive terminology, meanings, phrases, and modes of expression~\cite{mellinkoff2004}. This implies that conventional semantic analysis may fail to capture the correct or intended meaning of legal text.

\subsubsection{Intentional ambiguity}
Regulatory norms are by nature ambiguous and abstract to a certain degree~\cite{breaux2007}. Previous regulatory RE research has considered ambiguity as related to the necessity to consider additional regulatory sources~\cite{ghaisas2018} or alternative meanings of words. From a legal point of view, legal rules demand situational context~\cite{wroblewski1985}. Hence, the semantic analysis of regulatory norms alone cannot result in regulatory requirements---the situational context must be considered too. This was partially highlighted in previous research. For example, Otto et al.~\cite{otto2009} pointed to the necessity of taking the context of a software system into account. Furthermore, Jureta et al.~\cite{jureta2013} emphasized that regulatory RE focuses on a particular interpretation of the law in the context of software development. However, they do not suggest a cohesive approach to address ambiguity by considering the systems context. 

\subsubsection{Legal interpretation as a basis for the application of regulatory norms}
Regulatory norms are applied using legal interpretation or 'pragmatic interpretation' as determined in~\cite{dascal1988}. The main goal of legal interpretation is to ``\textit{give a meaning to a legal text}'' and answer the question ``\textit{what meaning to attach to the text}''~\cite{barak2011} in a given situation~\cite{dascal1988}. Thus, legal interpretation is different from the semantic analysis in its nature as its main purpose is to identify and assign a meaning relevant in a particular context rather than to obtain the meaning from the regulatory text alone. Some of the previous research has specifically focused on legal interpretation (e.g., ~\cite{muthury2016, ghanavati2015}). Still, these studies focused mainly on interpretation of regulatory norms per se, and ignored a more detailed consideration of the process of assignment of meaning in a particular context.

\subsection{Non-structured approach regulation syntax}\label{sec:syntax}
Previous research in linguistics found that the language of law is characterized by specific rules of entailment, dependent on its syntactic structures~\cite{kaplan1993}. Previous regulatory RE research has paid attention to concepts that can be used for the semantic analysis of regulations (e.g., as a part of an upper-ontology~\cite{breaux2007} or as a system of semantic metadata concepts~\cite{sleimi2021}). It is widely accepted that regulatory norms are expressed in the form of rights and obligations~\cite{sleimi2021}. In our opinion, rights and obligations can also be considered as an element type of the structure of regulations or, in other words, regulatory syntactic categories. Multiple RE methods using certain concepts for the analysis of regulatory norms are based on the Hohfeldian legal theory~\cite{hohfeld1913}, developed for judicial reasoning more than 100 years ago. While concepts from this legal theory are often reused as is, they are usually amended with other concepts derived in unclear or purely argumentative ways without taking legal knowledge into account. Such new concepts are usually also not defined by the RE perspective on regulatory requirements. For example, many regulatory RE approaches suggest the application of the concept of ``actor.'' An analysis of GDPR (Regulation (EU) 2016/679), for example, with the application of such an approach can result in the identification of the actors ``data subject'' and ``supervisory authority.'' Still these 'actors' fulfil different functions from the regulatory perspective. Similarly, the concept of an ``actor'' does not directly map to the RE concept of an ``user.'' The concept of ``actor'' has no equivalent meaning in neither regulations nor requirements.

\subsection{Reductionist approach to regulatory norms}\label{sec:redux}
While many researchers emphasized the complex nature of regulations, many approaches that had the ambition to address the legal knowledge aspects of regulatory norms have achieved it to a limited degree. Such approaches usually relied on reductionist perspective that omit important aspects of legal knowledge and intentionally concentrate on one or few of facets of legal knowledge and perspective~\cite{boella2014}. For example, regulatory RE methods apply an exhaustive system of concepts for semantic analysis that cannot cover elements of different regulations. Many regulatory RE methods apply methods that either focus on the intent of the law \emph{or} its structure, but do not consider both~\cite{akhigbe2019}. Overemphasizing one facet of regulations, and using it as the only relevant perspective disregards the real regulatory intentions of a law. For example, goal modeling is intended to model regulatory norms as goals; nevertheless, regulatory goals are often formulated directly by regulators in special norms and preambles to regulations. It is unclear how these original (explicit) goals, formulated by regulators and goals, derived from regulations (implicit), compare. Such reductionism results in the perception of regulatory RE methods as naive and relatively simplistic from the perspective of legal practice~\cite{boella2014}.

The unsolved problems in regulatory RE research are closely interrelated and interlinked. The absence of a systematic approach to system, structure, and semantics of regulations creates potential inconsistencies and results in methods that promote reductionism and oversimplification in regulatory RE methods. In the following section, we outline a systematic approach to address these issues.

\section{Practice-oriented regulatory RE}\label{sec:concept}
\begin{figure*}[htp]
\centerline{\includegraphics[width=0.85\linewidth]{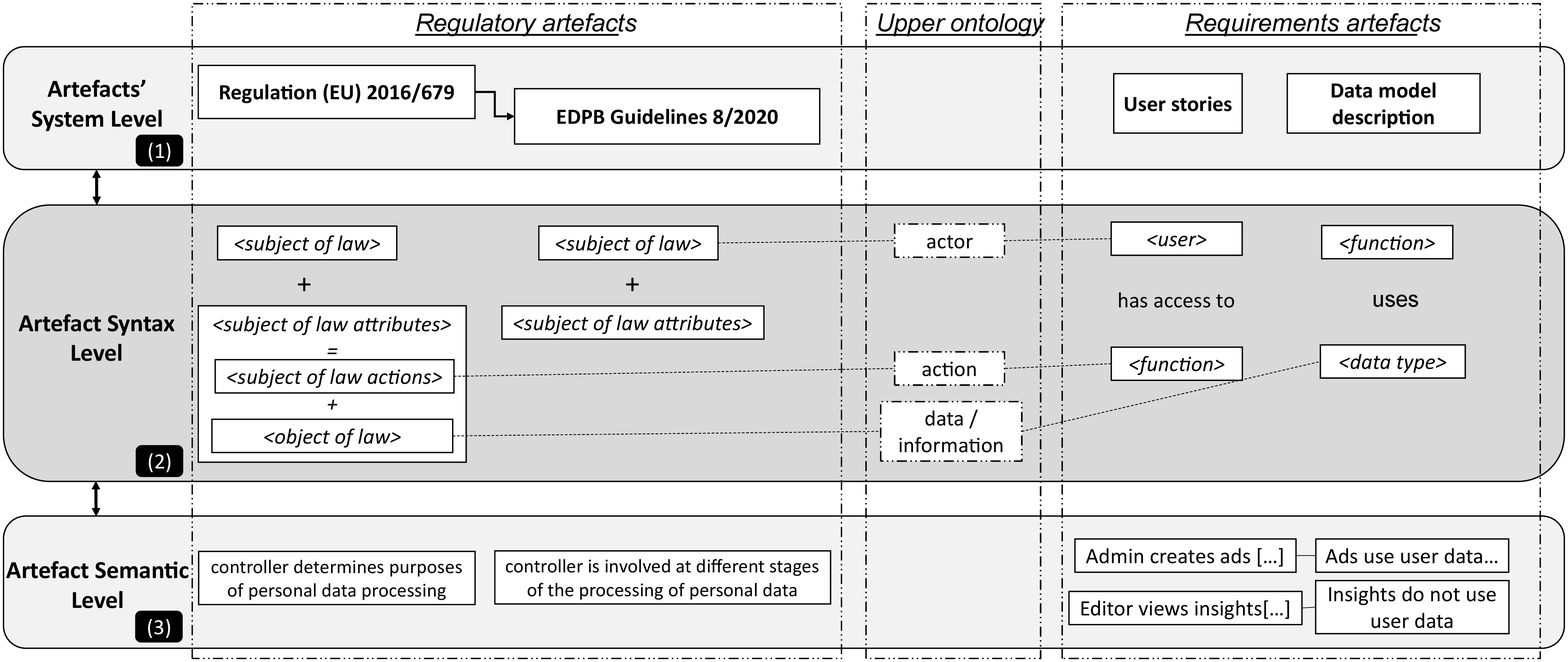}}
\caption{Illustration of the application of (1) artefact orientation for processing of artefacts and (2) legal interrelation methodological framework for interrelation of regulations and requirements}
\label{fig:example}
\end{figure*}

We suggest to address the challenges introduced in Section~\ref{sec:wicked} by resolving the research gaps elaborated in Section~\ref{sec:gaps} with an approach based on three pillars: (1) adaptation of the concept of legal interpretation to RE concepts, coupled with (2) artefact-orientation as an approach for performing interpretation, and (3) extension of the existing state-of-the-art Artefact Model for Domain-independent Requirements Engineering (AMDiRE)~\cite{fernandez2013} model with the elements required for regulatory RE purposes. We describe these three pillars next and conclude this section by outlining our empirical research strategy.

\subsection{Legal interpretation driving regulatory RE}\label{sec:interpretation}
We borrow the concept of legal interpretation from the legal domain as a basic method to elicit regulatory requirements from regulatory norms. Legal interpretation is directed at attaching meaning to regulatory norms in the context of a particular software development project. Attachment of meaning to regulatory norms and the contextuality of legal interpretation imposes two major principles for regulatory RE: (1) models of regulatory norms on one side and requirements on the other side which are independent of each other, and (2) the interrelation of legal norms and requirements to attach meaning and elicit regulatory requirements from them. 

The legal interpretation approach allows us to conceptually structure the regulatory RE process by treating software requirements as the context for legal interpretation of regulatory norms. Such an approach requires that both regulatory norms and software requirements are modeled in a comprehensive way (addressing all major concerns and features) before they can be interrelated. To that end, we suggest that modelling of regulations needs to be conducted in an independent way---i.e., regulations need to be modeled by legal experts and software requirements by software engineers---on the basis of a common modelling framework. Such approach removes reductionism inherent to multiple regulatory RE methods (see Section~\ref{sec:redux}) as models will contain all the elements and complexity deemed relevant by the corresponding domain experts. Additionally, independent modelling needs to address the problem of legal experts' involvement, ensuring better acceptability and maintainability of such models (see Section~\ref{sec:wicked}). Our main motivation for the application of independent modeling is to enable empirical research described further in Section~\ref{sec:empiricism}.
Flexibility in modelling comes with the trade-off of requiring an approach that interlinks regulatory and requirements models. We outline our approach next.

\subsection{Artefact-oriented modeling}\label{sec:ao}
As a basis for modelling both software requirements and regulations, we suggest an artefact-oriented approach. We use this approach to model basic elements of the analysis in our approach --- regulatory and requirements artefacts. Artefact orientation can be effectively used to represent software requirements as a system of interconnected deliverables (e.g., documents) containing the information necessary during the RE process~\cite{fernandez2019}. Artefact orientation in software engineering is based on the philosophical concept of artefact as an object created for a certain purpose~\cite{hilpinen1992} which makes it applicable to regulations. The interest in artefact orientation has grown in recent years particularly for the application of the general artefact philosophy in legal research~\cite{burazin2016}.
We use a running example centered on European privacy law to illustrate the use of artifact-orientation for regulatory RE.

Figure~\ref{fig:example} shows the artefact orientation as applied to regulatory and software requirements artefacts. Artefact-orientation suggests three main levels of artefact perception that reflect main levels on which artefacts are specified and perceived: (1) physical representation and system of artefacts, (2) syntax, (3) semantics~\cite{fernandez2019}. 

On the left side of Figure~\ref{fig:example}, at the system level (1) of regulatory artefacts, legal experts encode their knowledge by modeling interrelations between regulatory artefacts (e.g., identifying EDPB guidelines as one of the sources of regulatory norms) and aspects related to such interrelations (e.g., higher hierarchical position of GDPR, nature of cross-references between the artefacts).

In the context of regulations, we identify a regulatory artefact as any type of official mandatory normative document created for regulatory purposes. The three levels of artefact perception, introduced earlier, can also be applied in the context of regulations. This allows for a more systematic analysis of regulatory norms (see Section~\ref{sec:non-integrated}) as all three layers of artefact perception can be considered in their interconnection.

Legal knowledge is also integrated into the model on the artefact syntax level (see Figure~\ref{fig:example}, (2)). We define the syntax of regulatory artefacts as a set of rules that govern the structure of regulatory artefacts. Therefore, regulatory syntactic categories are types of information, information patterns and concepts that regulators use to specify regulations. Such regulatory syntactic categories are known to legal experts because they are used in regulatory practice and formulated in legal theories (e.g., in the Hohfeldian theory). Thus, such syntactic categories can be annotated by legal experts in regulatory texts (artefact semantic level, (3)), embedding  knowledge on what is relevant from the legal perspective and use wording consistent with the legal domain. 

In Figure~\ref{fig:example} artefact semantic level (3) represents not only the text itself but also the meaning intended by regulators and underpinned by system (1) and syntax (2) of a regulatory artefact. Coming back to our running example, the annotation of two excerpts from the regulatory norms (artefact semantic level (3)) results in the identification of four main syntactic categories (artefact syntax level (2)) in Figure~\ref{fig:example}: \texttt{<subject of law>}, \texttt{<subject of law attributes>}, \texttt{<subject actions>}, and \texttt{<object of law>}. Legal experts annotate them with their knowledge that in European privacy law ``controller'' is one of \texttt{<subjects of law>} (i.e., persons or entities that are empowered to realize rights and obligations), and ``personal data'' is an \texttt{<object of law>} (i.e., object or relations that are a concern for the law). As, to the best of our knowledge, there is no ontology of syntactic categories, we plan to develop one to capture such interrelations (e.g., that \texttt{<subject of law attributes>} (properties of the subject defined by the regulation) and \texttt{<subjects actions> (actions that the subject can take according to the regulation)} are related to \texttt{<subject of law>}).

We conjecture that syntactic categories help to make legal domain knowledge explicit. Although the interrelation of regulations and requirements need to be on the syntax level (2), we aim to establish an understanding of regulations on the semantic level (3). To that end, we suggest to establish the interrelation between regulatory and requirements syntactic categories by using an upper ontology. We identify regulatory syntactic category \texttt{<subject of law>} as belonging to the \texttt{[actor]} concept in the upper ontology. In parallel, we establish that \texttt{<user>} (as requirements syntactic category) also belongs to the \texttt{[actor]} concept. Accordingly, we can assume that ``controller'' (annotated as \texttt{<subject of law>}) can be interrelated to ``Admin'', ``Editor'' (belonging to \texttt{<user>} requirements syntactic category). The upper ontology allows us also to establish the interrelations between \texttt{<subject of law actions>}, \texttt{<object of law>} and \texttt{<user functionality>}, \texttt{<data type>}. 

Moving in Figure~\ref{fig:example} from the interrelated syntactic categories (2) to the semantic level (3), we can identify the exact meaning for both regulations and requirements. On the regulatory artefact side, the \texttt{<subject of law>} 'controller' + \texttt{<subject of law actions>} 'determines the purposes and means of the processing of' + \texttt{<subject of law actions>} 'may be involved at different stages of the processing of' + \texttt{<object of law>} 'personal data'. On the requirements side, the \texttt{<user>} 'Admin' + \texttt{<user functionality>} 'create ads, promotions' [that use] + \texttt{<data type>} 'user data' and \texttt{<user>} 'Editor' + \texttt{<user functionality>} 'view insights [that don't use]' + \texttt{<data type>} 'user data'. Analyzing the semantics of interrelated syntactic categories, we can reason about all of them except on 'personal data' on the regulatory side and 'user data' on the requirements side. It is not clear if they are equivalent. To clarify this, we need to shift to the syntactic level and identify annotated \texttt{<object of law attributes>} linked to 'personal data'. This generates a list of properties of personal data (e.g., 'information relating to an identified or identifiable natural person') that are required to interrelate 'personal data' and 'user data'. If such properties are specified in existing requirements they need to be identified. If these properties are absent, they need to be specified. In this way, the completeness of requirements needed for regulatory compliance can be evaluated and eventually achieved. To summarize, an artifact-oriented approach supports regulatory RE tasks by:
\begin{itemize}[noitemsep,topsep=0pt]
    \item encoding the main domain knowledge elements into syntactic categories, making their interpretation explicit.
    \item supporting regulation-centric RE that is not reductionist and follows regulatory reasoning.
    \item supporting the completeness of requirements specifications.
    \item embedding domain knowledge, enabling better communication between legal experts and requirements engineers.
\end{itemize}

\subsection{Extension of AMDiRE to AMDiRRE}\label{sec:amdirre}
\begin{figure}[htp]
\centerline{\includegraphics[width=0.85\linewidth]{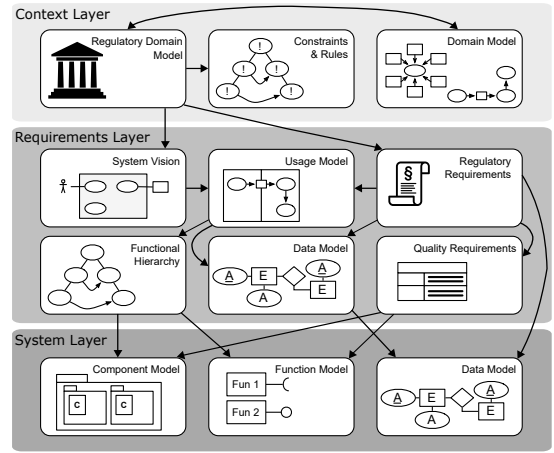}}
\caption{AMDiRRE with the new content items (Regulatory Domain Model, Regulatory Requirements) and relations}
\label{fig:amdirre}
\end{figure}

After the independent modeling of both regulatory and requirements artefacts and interrelating them, we extend the AMDiRE model as shown in Figure~\ref{fig:amdirre} with two new content items. AMDiRE serves as a checklist for the different aspects to be considered and assists to structure the specific RE tasks. AMDiRE contains content items that are logical groupings of the artefacts’ contents according to the information needed in the RE process~\cite{fernandez2013}. AMDiRE content items are organized on three layers of abstraction (Context Layer, Requirements Layer, and System Layer) that represent three main perspectives on modeling of a software system. The extended model - Artefact Model for Domain-independent Regulatory Requirements Engineering (AMDiRRE) - considers the system of requirements artefacts and integrates the models of regulatory artefacts in it. The first content item we add is the 'Regulatory Domain Model' that contains the artefact-oriented model developed by legal experts and is added to the Context Layer of the model. The second content item is 'regulatory requirements' item that contains (1) the results of the interrelation between regulatory domain model and requirements artefacts and (2) implications of such interrelations. For example, if 'user data is personal data', then 'user is data subject', 'personal data must be encrypted' etc. This way the 'Regulatory requirements' content item in Requirements Layer should provide an overview of new requirements or needed requirements changes combined with the reasoning for such changes.
The new content items are integrated with existing AMDiRE content items through dependencies and relations that encompass the main artefacts and information that regulatory requirements depend on or are related to.

\subsection{Empirical research strategy}\label{sec:empiricism}
In order to realize the approach outlined in this section, we need to answer a series of research questions. AMDiRRE relies on the identification of syntactic categories that enable regulatory and legal interpretation. Hence, our first research question is:
\begin{description}[noitemsep,topsep=0pt]
    \item[RQ1] What are the legal syntactic categories, concepts and theories relevant for the regulatory RE purposes?
\end{description}

In order to answer this question, we plan to conduct interdisciplinary qualitative analysis of regulatory documents in cooperation with legal researchers and experts to identify and annotate relevant categories and concepts. The research will be supported by an interdisciplinary literature review and will result in an ontology of regulatory syntactic categories.
Next, we plan to answer the following three questions:
\begin{description}[noitemsep,topsep=0pt]
    \item[RQ2] To what extent can the developed ontology be used to interrelate regulatory and requirements syntactic categories?
    \item[RQ3] To what extent can the suggested approach enable communication between legal experts and requirements engineers?
    \item[RQ4] How resilient is the legal interpretation methodological framework against regulatory variability?
\end{description}

We plan to conduct an interdisciplinary case study involving requirements engineers and legal experts. We will apply the artefact-oriented legal interpretation methodological framework and AMDiRRE supported by the ontology of regulatory syntactic categories in real life settings. We expect to verify (1) methodological framework in conjunction with AMDiRRE and the developed ontology, (2) applicability of the approach to facilitate interaction between legal experts and requirements engineers, (3) resilience against regulatory variability.
As the result of both empirical enquiries we plan to investigate the feasibility of automation of syntactic analysis of regulations and develop an empirical strategy for further automation based on text mining as basic method to support both legal experts and requirements engineers in applying our approach.

\section{Regulatory RE: 10 years from now}\label{sec:vision}
Already today, law makers are lagging far behind the pace of technological development. There is little evidence that problems of overregulation and 'technology-unfriendly' regulations will be resolved on the regulatory side of the equation alone. We conjecture that in the coming decade, regulatory compliance of software systems will become a concern on even wider scale. We see regulatory RE as one of the main fields contributing to resolution of software compliance challenges in an all-encompassing way satisfying regulatory goals, software business needs and software users' concerns. Next we outline some of the compliance related needs that are crucial in the years to come, together with the expected contribution of our suggested approach. Future regulatory RE should:

\emph{Empower} software engineers to consider regulatory compliance and related risks already at the early stages of software system development. In our approach this is enabled by its applicability to early stage software development artefacts (e.g., business plans).

\emph{Reduce} the burden on financial and human resources in software projects. We suggest to achieve it through (1) making involvement of legal experts more effective, (2) application of lightweight artefact-oriented modelling.

\emph{Help} requirements engineers to better understand and navigate regulations to ensure completeness, validity and adaptivity in compliance. We strive to assure that using independent and comprehensive regulatory modelling, and automation.

\emph{Assist} supervisory agencies to conduct systematic and effective verification of regulatory compliance.

\section{Conclusion}\label{sec:conclusion}
Regulatory RE research faces comprehensive interdisciplinary challenges such as the specific nature of regulatory requirements and legal knowledge. Existing research gaps in regulatory RE prevent a systematic approach to solve these challenges. Empirical research has a special role in resolving existing challenges and tackling potential future hurdles, due to the potential to better capture interdisciplinary phenomena. To achieve this, we suggest our approach based on the concept of legal interpretation combined with artefact-orientation and artefact modeling that relies on empirical methods.

\begin{acks}
We would like to acknowledge that this work was supported by the bidt through the Coding Public Value project and the KKS foundation through the S.E.R.T. Research Profile project at Blekinge Institute of Technology.
\end{acks}

\bibliographystyle{ACM-Reference-Format}
\bibliography{references}
\end{document}